\documentclass[aps,pra,twocolumn]{revtex4-1}
\usepackage{graphicx} % uncomment this line to include the graphicx package
\usepackage{tabularx}
\def\gapx{\lower 2pt \hbox{$\buildrel>\over{\scriptstyle{\sim}}$\ }}
\def\lapx{\lower 2pt \hbox{$\buildrel<\over{\scriptstyle{\sim}}$\ }}
\def\ph2{{\it p}-H$_2$}
\def\beq{\begin{equation}}
\def\eeq{\end{equation}}
\def\Am3{\AA$^{-3}$}
\begin{document}

\widetext
%\draft
\title{Atomic displacements in quantum crystals}
\author{Marisa Dusseault$^\star$ and Massimo Boninsegni} 
\affiliation{Department of Physics, University of Alberta, Edmonton, Alberta, Canada T6G 2G7}
\date{\today}

\begin{abstract}
Displacements of atoms and molecules away from lattice sites in  helium and parahydrogen solids at low temperature 
have been studied  by means of Quantum Monte Carlo simulations. 
In the  bcc phases of $^3$He and $^4$He, atomic displacements are largely 
quantum-mechanical in character, even at melting. The computed Lindemann ratio at melting is found to be in good agreement with experimental results for $^4$He. Unlike the case of helium, in solid parahydrogen 
there exists near melting a significant thermal contribution to molecular vibrations, accounting for roughly half of the total effect. Although the Lindemann ratio at melting is in quantitative agreement with experiment,  computed molecular mean square fluctuations feature a clear temperature dependence,  in disagreement with recent experimental observations.
\end{abstract}
%\pacs{67.10.Hk, 61.20.Ja, 67.63.Cd,  67.}
\maketitle
%\narrowtext
% \begin{multicols}{2}

\section{Introduction}

The denomination ``quantum solid'' refers to crystals displaying 
marked quantum-mechanical effects \cite{gov}. These include a kinetic energy per particle significantly above its classical value 3$T$/2 \cite{note0}, as well as considerably more pronounced particle excursions away from their equilibrium (lattice) positions than observed in most solids. The latter assertion can be phrased quantitativey in terms of  the 
so-called Lindemann ratio \cite{lindemann}  $\delta=u_{rms}/a$,  $u_{rms}$ being the root-mean-square displacement  of particles away from lattice sites, and $a$  the distance between nearest-neighboring sites. This ratio increases with temperature, and in most substances takes on a value close to 0.1 at melting; however,  in a highly quantal solid such as $^4$He it 
can be as high as $\sim 0.3$, even at temperature $T$=0, as a result of the large zero-point motion resulting from the light mass of the atoms. 
Indeed, for a long time the most commonly accepted explanation for the failure of He to solidify at low temperature, under the presure of its own vapour, has been that 
the large zero-point motion of its light atoms acts  to destabilize the crystal phase (this belief has been recently challenged \cite{munich12}). An interesting theoretical question remains that of the role played by quantum mechanics in the melting of solids made of light atoms or molecules.
\\ \indent Helium is, of course, the archetypal quantum solid \cite{henry} but sizable quantum effects have also been predicted and observed for solid parahydrogen ({\it p}-H$_2$). Although its constituent particles are molecules of mass one 
half of that of a $^4$He atom, {\it p}-H$_2$  is nonetheless a stronger crystal than helium at comparable thermodynamic conditions, owing to the depth of the attractive well of the intermolecular potential, roughly three times that between two helium atoms. For this reason, unlike helium parahydrogen is a crystal at $T$=0 (even in reduced dimensions \cite{me1,me2}), and in many respects the behaviour of solid  {\it p}-H$_2$  interpolates between that of a classical crystal and solid helium.  It has been recently suggested, however, that the melting of a crystal of \ph2 may be driven primarily by quantum-mechanical effects \cite{alonso}.
\\ \indent
Both the mean kinetic energy per particle, as well as the mean square displacement $\langle u^2\rangle \equiv u_{rms}^2$ around lattice sites, can be accessed experimentally by either x-ray or neutron scattering \cite{bennington}; in particular, estimates of $\langle u^2\rangle$  can be obtained through the measurement of the Debye-Waller factor \cite{simmons,simmons2}. On the theoretical front, first principle \cite{notefp} calculations of both quantities can be carried out by means of Quantum Monte Carlo (QMC) simulations; thus, a comparison of theoretical results with experimental data represents a cogent test of the current theoretical understanding of the physics of the simplest quantum crystals,  as well as of the most commonly adopted microscopic models thereof. 
\\ \indent
The most important theoretical assumptions built into the vast majority of QMC calculations are {\em a}) the adoption of a pair potential to describe the interaction of atoms or molecules  {\em b}) the neglect of quantum-mechanical exchanges of indistinguishable particles. 
The use of pair potentials has been shown to afford a rather accurate theoretical description of energetic and structural properties of solid He \cite{whitlock,herrero2} (even though  the inclusion of  three-body terms improves significantly the agreement with the experimental equation of state \cite{whitlock}), 
as  well as  \ph2 \cite{operetto}. Exchanges of indistinguishable particles are almost invariably neglected in most  theoretical calculations of 
crystal properties, the main justification being the infrequency with which such exchanges occur \cite{guyer,richards,sg}, even in a highly quantal crystal like $^4$He. 
\\ \indent
It has been suggested, however, that quantum-mechanical exchanges may play a {\it major} role in the stabilization of the bcc phase of solid $^3$He. A significant test of this hypothesis would consist of a comparison of computed and measured values of $\langle u^2\rangle$ in this system, as one of the most important consequences of quantum-mechanical exchanges is a much greater atomic mobility, resulting in more pronounced fluctuations around lattice sites than if exchanges were absent \cite{pederiva}.
\\ \indent
A detailed comparison of experimental and theoretical results for $\langle u^2\rangle$  has been carried out for fcc and hcp  $^3$He and $^4$He, at temperatures between 5 K and 35 K, yielding satisfactory agreement \cite{draeger,simmons}. To our knowledge, no microscopic calculations have been performed for the bcc phase of the two isotopes, which occurs at considerably lower temperatures and coexists with fluid phases whose physics is dominated by quantum exchanges. On the experimental side, we are only aware of measurements of the Lindemann ratio in bcc $^4$He.\\ \indent
In this paper we report results of QMC calculations of the mean square atomic excursions away from lattice positions in bcc $^4$He and $^3$He at the melting temperatures $T$=1.6 K ($^4$He) and 0.65 K ($^3$He). We also computed the same quantity for the equilibrium  hcp phase of \ph2  from a low temperature $T$=1 K, all the way to the melting temperature $T$=13.8 K. In all of our calculations pair potentials are used and quantum-mechanical exchanges neglected. We compute separately thermal and quantum-mechanical contributions to atomic and molecular displacements.
\\ \indent
Computed values of $\langle u^2\rangle$ are found to be in excellent agreement with experiment for bcc $^4$He; for bcc $^3$He our estimated value of the Lindemann ratio $\delta$ is  $\sim$ 6.5\% lower than existing theoretical estimates.  In principle, there is no reason not to expect comparable accuracy for both isotopes. However, if particle exchanges are as prominent in this phase as suggested in Ref. \cite{pederiva},  it is possible that one may see a difference between the experimentally measured value and one predicted by a calculation neglecting atomic exchanges, as these are  likely to result in a significant enhancement of the mobility of atoms in the crystal at melting.\\
\indent
The results for $\langle u^2\rangle$ in hcp \ph2  show a clear dependence on temperature, of a magnitude that appears to be experimentally observable; this is in qualitative and quantitative disagreement with recent measurements. We find that, while atomic excursions away from lattice sites in bcc He  are largely quantum-mechanical in character, in \ph2 at melting thermal and quantum-mechanical contributions are very nearly equal in magnitude. Thus, according to this criterion solid \ph2 may be regarded as neither  completely ``quantum'' nor ``classical'', but rather as a system sitting on the dividing line between the two physical behaviours.
\\ \indent
We also report in this paper our estimates for the pressure and/or the  mean kinetic energy per particle, and compare them with available experimental determinations.
The remainder of this article is organized as follows: in Sec. \ref{model} we introduce the microscopic model and offer details of the calculation carried out in this work; in sec. \ref{res} we illustrate our results, outlining our conclusion in sec. \ref{conclusions}.

\section{Model}\label{model}
Consistently with most theoretical studies, our systems of interest, namely solid helium and parahydrogen, are  modeled as ensembles of $N$ point-like particles 
of spin zero,
enclosed in a parallelepipedal cell of volume $\Omega$ with periodic boundary conditions.
The quantum-mechanical many-body Hamiltonian is the following:
\begin{equation}\label{one}
\hat H = -\lambda\sum_{i=1}^N \nabla_i^2 + \sum_{i<j} V(r_{ij}) 
\end{equation}
Here, $\lambda$=6.0596  K\AA$^2$ for $^4$He, 8.0417 K\AA$^2$ for $^3$He and  
12.031 K\AA$^2$ for \ph2, while  
$V$ is the potential describing the interaction between two particles, atoms or molecules, assumed here to depend only on their relative distance. The results
 presented here for the helium isotopes were obtained using the Aziz pair potential \cite{aziz}, whereas the Silvera-Goldman  
potential \cite{SG} was used for \ph2. It is worth mentioning that these are not the only potentials that have been used in previous simulation work, but it seems fair to state that they are the most commonly adopted. They have both been shown to provide rather accurate quantitative descriptions of the thermodynamics of the solid phase of helium \cite{kalos} and \ph2 \cite{herrero,operetto}.
Naturally, in principle a more accurate model would go beyond the simple pair decomposition, including, for instance, interactions among triplets; however, published numerical work (e.g., on helium) has given strong indications that three-body corrections, while significantly affecting the estimation of the pressure, have a relatively small effect on the structure and dynamics of the system, of interest here \cite{whitlock,bpc,syc}.
\section{Methodology}
The thermodynamic properties of the system, as modeled by the many-body Hamiltonian (\ref{one}), were studied by means of  numerical simulations, based on the continuous-space Worm Algorithm (WA) \cite{worm,worm2}.
This is a fairly well-established (Monte Carlo) methodology, based on R. P. Feynman's path integral formulation of quantum statistical mechanics \cite{feynman}. It allows one to obtain essentially exact numerical estimates of thermodynamic properties of quantum many-body systems at finite temperature, directly from the microscopic Hamiltonian.  The reader is referred to Ref. \onlinecite{worm2} for a thorough description of  
the technique.  The specific implementation 
utilized in this project is {\it canonical}, i.e., we keep the number $N$ 
of particles fixed \cite{fabio}.  Other technical aspects of the calculations 
are common to any other  QMC simulation scheme.  The usual fourth-order high-temperature propagator was adopted here \cite{jltp2}; convergence of 
the estimates was observed for a time step 
$\tau= 3.125\times 10^{-3}$ K$^{-1}$, for both He and \ph2. 
Because the computational cost was negligible, all estimates reported 
here were obtained using twice as small a time step, in order to be on 
the safe side. 
\\ \indent
As mentioned above, the sampling of quantum-mechanical exchanges of identical particle, in principle allowed by the methodology utilized here, is explicitly {\rm excluded}, i.e., we treat particles as {\em distinguishable}. This is an excellent assumption for \ph2, as well as for $^4$He  in the temperature range explored here. On the other hand,  for $^3$He at the temperature of interest exchanges do occur, albeit still relatively rarely. The exclusion of exchanges is deliberate, as we are interested in assessing their possible importance through a comparison with experimental measurements; however, it ought to be reminded that inclusion of exchanges in a Fermi system like $^3$He would be rendered much more complicated by the appearance of the well-known ``sign" problem, plaguing any current QMC methodology. In the absence of exchanges, the WA is very similar to conventional Path Integral Monte Carlo \cite{jltp2}, although the presence of an open world line (path) in the WA allows for direct  computation of the one-body density matrix \cite{bon09}, for which however we do not report results here.
\\ \indent
The calculations of the kinetic energy per particle and of the pressure, as well as of structural quantities such as the pair correlation function, are standard \cite{jltp2}; on the other hand, that of $\langle u^2\rangle$ has been performed relatively infrequently so far, and for this reason we offer here a few relevant details, with the general remark that we largely duplicate the procedure illustrated in Ref. \onlinecite{draeger}. \\ \indent
The mean square displacement of a particle away from a lattice site $\langle u^2\rangle$ in a crystal at temperature $T$ is defined as
\beq\label{usq}
\langle u^2\rangle = \frac{T}{N} \ \sum_{i=1}^N \biggl \langle \int_0^{1/T} du\ \biggl ({\bf r}_i(u)-{\bf b}_i \biggr )^2\biggr \rangle
\eeq
where the sum runs over all the atoms or molecules in the system, $\{{\bf b}_i\}$ is a set of  fixed lattice positions consistent with the crystal structure of interest, $\langle ... \rangle$ stands for quantum-mechanical thermal average and the integral is over the imaginary-time path ${\bf r}_i(u)$ of the $i$th  particle \cite{feynman}. 
It is possible to identify two separate contributions to $\langle u^2\rangle$ in (\ref{usq}), namely \cite{gillan,herrero} 
\beq\label{break}
\langle u^2\rangle = \langle u^2\rangle_Q+\langle u^2\rangle_T
\eeq
where
\beq\label{qq}
\langle u^2\rangle_Q = \frac{T}{N} \ \sum_{i=1}^N \biggl \langle \int_0^{1/T} du\ \biggl ({\bf r}_i(u)-{\bf {\bar r}}_i \biggr )^2\biggr \rangle
\eeq
and
\beq\label{tt}
\langle u^2\rangle_T = \frac{1}{N} \ \sum_{i=1}^N \biggl \langle \biggl ({\bf b}_i-{\bf {\bar r}}_i \biggr )^2\biggr \rangle
\eeq
having defined the single particle path centroid as
\beq
{\bf {\bar r}}_i=T\ \int_0^{1/T} du\ {\bf r}_i(u)
\eeq
The usefulness of this breakdown lies in the fact that $\langle u^2\rangle$ is dominated by either $\langle u^2\rangle_Q$  or $\langle u^2\rangle_T$ in different temperature limits. Specifically, at high temperature (i.e., in the classical limit) $\langle u^2\rangle_Q$ becomes negligible and $\langle u^2\rangle\approx \langle u^2\rangle_T$, as the  imaginary-time path of a quantum particle shrinks to a point. On the other hand, in the $T\to 0$ limit it is $\langle u^2\rangle\approx\langle u^2\rangle_Q$, as particle excursions away from lattice sites are mostly quantum-mechanical in nature (i.e., zero-point motion). Thus, Eq. (\ref{break}) can be used to assess quantitatively the relative importance of quantum-mechanical versus classical (i.e., thermal) contributions to atomic/molecular displacements at a given $T$. In particular, we shall make use of the ratio $\chi \equiv \langle u^2\rangle_Q / \langle u^2\rangle$, which tends to zero at high $T$ and approaches unity as $T\to 0$.
\\ \indent
Implicit in the definitions (\ref{usq}) and (\ref{tt}) is the assumption that at each imaginary time along the paths, every particle remain close to one and the same lattice position.  In a typical simulation,  particles are initially placed at lattice sites ${\bf b}_i$, $i=1$ through $N$, but obviously wander away  in the course of a run; for the systems and thermodynamic conditions considered in this study, we observed that particles for the most part remain close to their initial lattice sites
\cite{noteaa}.
At low temperature, however, especially for the lighter He isotope, single-particle paths can become sufficiently extended in space that at different imaginary times a particle may be close to {\em different} lattice sites,  even in the absence of quantum-mechanical exchanges. In order to avoid complications, we discarded the contribution to  the accumulated statistical average of $\langle u^2\rangle_T$ of  such  (very rare) single-particle paths. Because the fraction of single-particle paths discarded has never exceeded $\sim 10^{-5}$ in any of our simulations, we believe this not to result in a significant bias of our estimates.
It may be noted that it  is roughly equivalent to the procedure 
implemented in Ref. \onlinecite{draeger}, consisting of rejecting sampling moves that take any particle further away from its assigned lattice site than a  
pre-defined``tether" distance, typically of the order of the lattice constant.
\\ \indent
The results presented here were obtained by 
simulating systems comprising a number of particles $N$ up to 1024 for He, 4096 for \ph2. As shown in Ref. \onlinecite{draeger}, estimates of $\langle u^2\rangle$ feature a nontrivial dependence on $N$, especially near melting. We come back to this point when discussing the results in detail.
\section{Results}
\label{res}
\subsection{Helium}
We begin by illustrating the results for the bcc phase of the two isotopes of He, at melting. We begin with $^4$He at $T=1.6$ K, and density $\rho=0.02854$ \AA$^{-3}$.
%------------
\begin{figure}[!t]
\includegraphics[width=6.0cm]{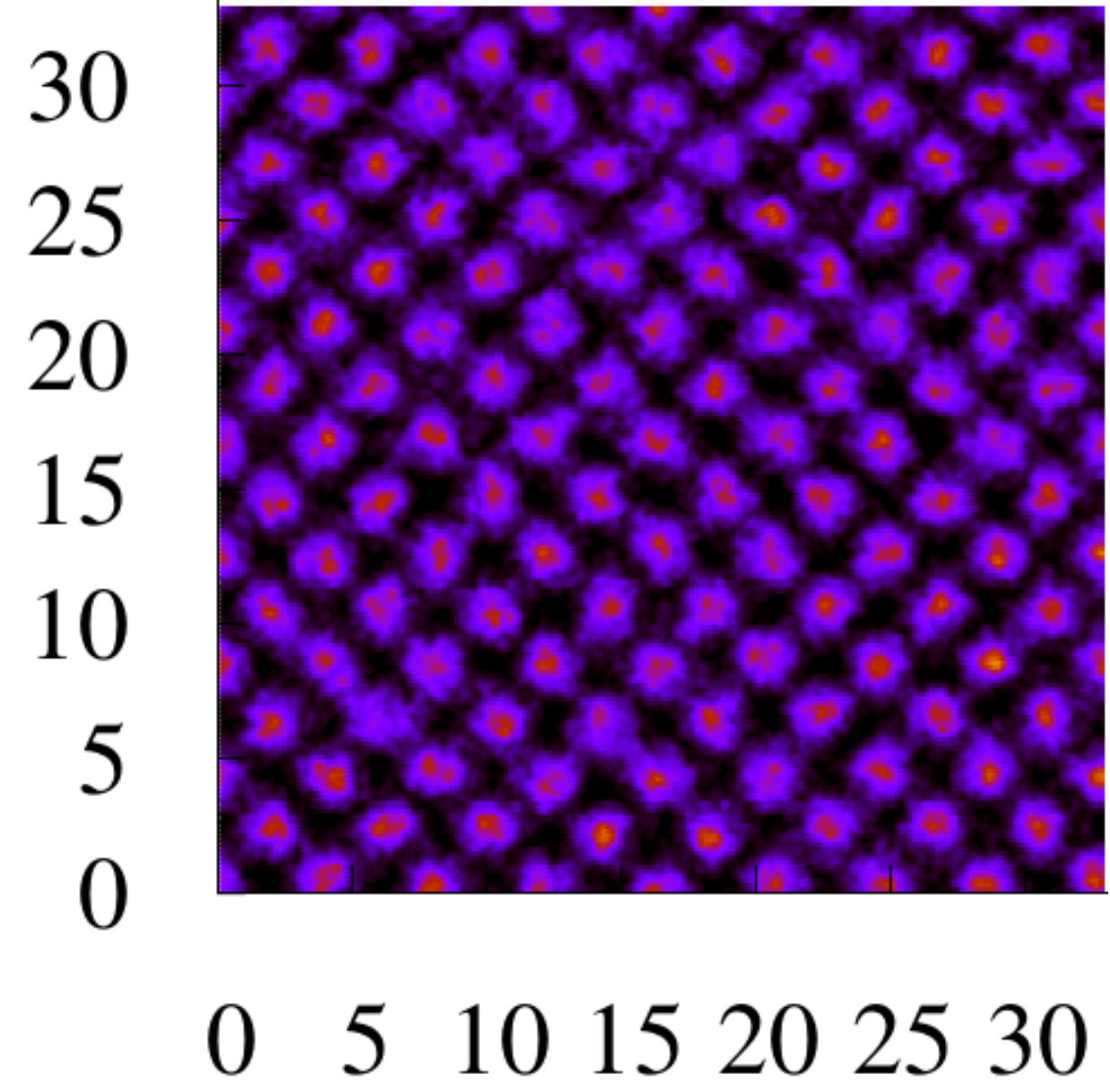} %{figure4}
\caption{{\em Color online.} Snapshot of a many-particle configuration (world lines) for bcc $^4$He, at temperature $T$=1.6 K, density $\rho=0.02854$ \AA$^{-3}$. View is along one of the three main (equivalent) crystallographic directions. The total number of particles is 1024. Only the traces of the 128 in two adjacent planes are visible.
Distances are in \AA.}
\label{f1}
\end{figure}
%---------------

\begin{figure}[!h]
\includegraphics[width=8.0cm]{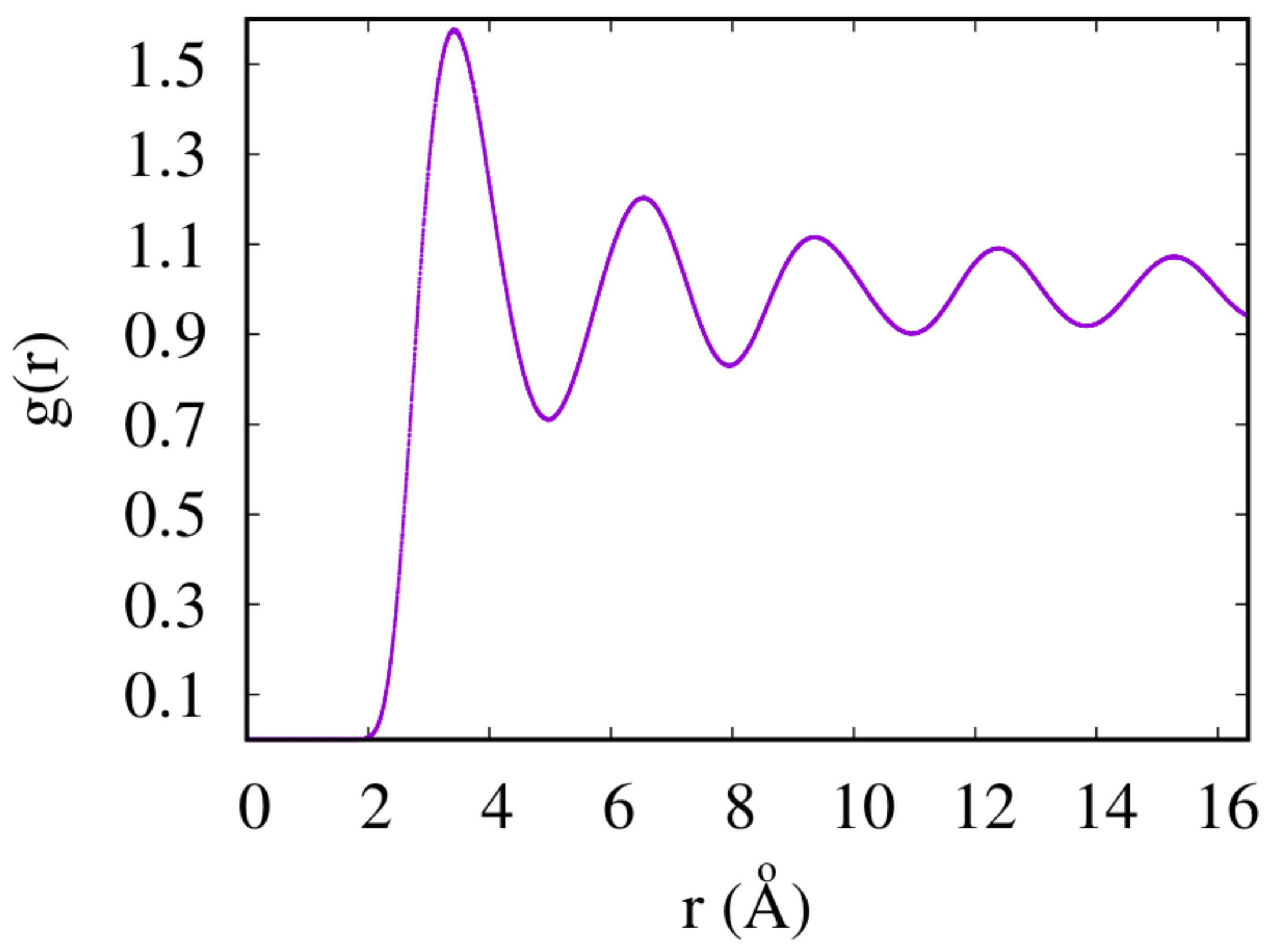} %{figure4}
\caption{{\em Color online.} Pair correlation function $g(r)$ for bcc $^4$He at $T$=1.6 K and density $\rho=0.02854$ \AA$^{-3}$, computed for a system of
$N$=1024 atoms.}
\label{f2}
\end{figure}

We simulated systems comprising $N$=128, 432 and 1024 particles. As we shall discuss below, the dependence on the system size of the averages of different physical observables is very different. Because of the proximity of the system to the fluid phase, we periodically inspected configurations in order to ensure that the system be still in the crystal phase. Fig. \ref{f1} shows a snapshot of a typical many-atom configuration generated in the course of a single run. The crystalline arrangement of the atoms is clear.
\\ \indent
The computed value of the pressure is 26.7(2) bars, which is in very good agreement with experiment \cite{hoffer,grilly}.
The kinetic energy per $^4$He atom $E_k$ is equal to 23.82(2) K in a system of $N$=128 atoms and remains unchanged, within statistical uncertainties, in systems  comprising more atoms (this is found to be the case also for $^3$He and \ph2). A direct comparison of this result  with available experimental and theoretical estimates is tricky because of small differences in the thermodynamic conditions. However, as shown in Table I, there appears to be overall consistency, also considering that the theoretical calculation of Ref. \onlinecite{rota} made use of slightly different interatomic potentials.
\begin{table}[h]\label{t1}
\centering
  \caption{Experimental and theoretical estimates of the kinetic energy per atom in the bcc phase of $^4$He.
Statistical uncertainties, in parentheses, are on the last digit.}
  \label{tab:table1}
  \begin{tabularx}{.48\textwidth}{lccccccccccccccccccccr}
\hline\hline\\
Method  & & &&&&& $T$(K)  & &&&&&& $\rho$ (\AA$^{-3}$) &&&& &&&$E_k$ (K)\\
    \hline\\
    This Work &&&&&&& 1.6 &&&&&&& 0.02854 &&&&&&& 23.82(2)\\ \\
Expt (Ref. \onlinecite{blasdell}) &&&&&&&1.725 &&&&&&&0.0288 &&&&&&&23.7(3)\\ \\
PIMC (Ref. \onlinecite{cep}) &&&&&&& 1.6 &&&&&&&0.0288 &&&&&&& 24.4\\ \\
PIMC (Ref. \onlinecite{rota}) &&&&&&&  1.5 &&&&&&&0.02857 &&&&&&& 23.936(5)\\ \\
\hline
  \end{tabularx}
\end{table}
\\ \indent
Fig. \ref{f2} shows the pair correlation function $g(r)$ computed for a system of $N=1024$ atoms. The curve features the characteristic, persistent oscillations of a crystalline system. Aside from the obvious fact that simulating a larger system allows one to access this quantity at greater distances, we observe small differences between the results obtained with the smaller system size simulated here (i.e., $N=128$) and those for $N=1024$, the height of the main peak approximately 1\% lower in the latter case.
%------------
%---------------
%------------
\begin{figure}[!h]
\includegraphics[width=8.0cm]{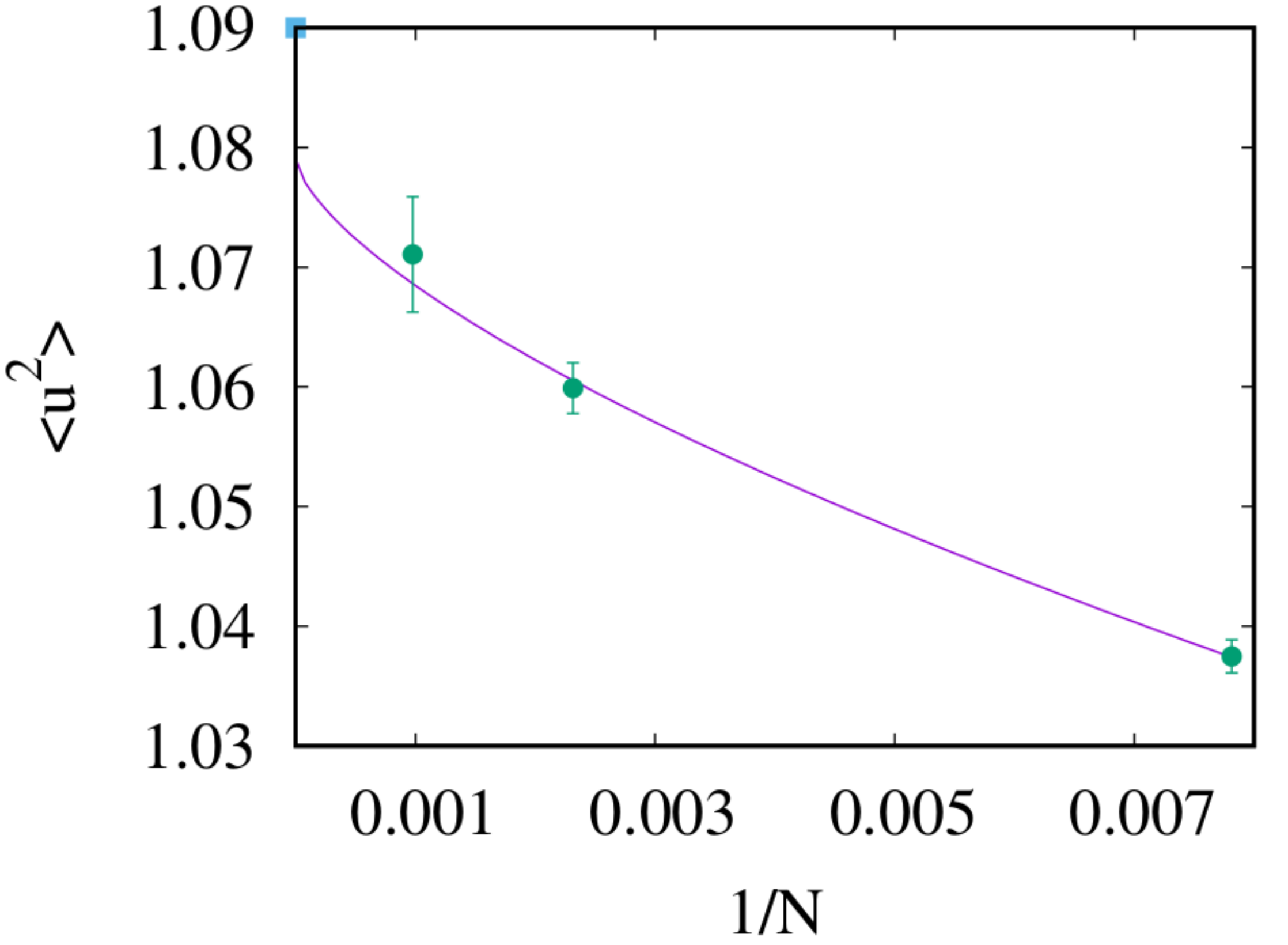} %{figure4}
\caption{{\em Color online.} Mean square atomic displacements (in \AA$^2$) away from lattice sites in bcc $^4$He at the
melting temperature $T$=1.6 K and density $\rho=0.02854$ \AA$^{-3}$. Shown here are the results for the three system sizes simulated (see text). Curve through the points is a fit to the data  based on the expression 
$\alpha+\beta N^{-2/3}$. The box at the left upper corner represents the experimental estimate from Ref. \onlinecite{henry}.}
\label{f3}
\end{figure}
%---------------
\\ \indent
Next we discuss our results for the mean square atomic displacements in bcc $^4$He. Our estimates of $\langle u^2\rangle$ are shown in Fig. \ref{f3} for the three system sizes considered here (see above). The first observation is that the ratio $\chi\approx 0.75$, i.e., atomic displacements away from lattice sites at melting are mainly quantum-mechanical in this crystal. Accordingly, based on the arguments illustrated in Ref. \onlinecite
{draeger}, we fit the data in Fig. \ref{f3} with the expression \beq\label{fitting}f(N)=\alpha+\beta N^{-\gamma}\eeq with $\gamma=2/3$ which yields an extrapolated value of $\langle u^2\rangle$ in the thermodynamic limit equal to 1.080(2) \AA$^2$ (the resulting Lindemann ratio is $\delta=0.291(1)$).
This is in remarkable agreement with the experimental value \cite{mink} of 1.09 \AA$^2$. 
Altogether, these results point to fairly good, quantitative agreement with experiment afforded by the relatively simple microscopic model of Eq. (\ref{one}), based on a pair potential.
\\
\indent
Let us now consider the bcc phase of the lighter isotope of helium, namely $^3$He, at $T$=0.65 K and at a density $\rho=0.02458$ \AA$^{-3}$, i.e., along the melting line. Considerably fewer experimental results have been published for $^3$He than for $^4$He, as neutron scattering measurements are complicated by the significant neutron absorption cross section.
\\
\indent
In this case, the value of the pressure as it emerges from the calculation is 31.8(2) bars, in excellent agreement with experiment \cite{mills}, just like for $^4$He. 
To our knowledge no estimates of the atomic kinetic energy have been reported at the thermodynamic conditions considered here. Indeed, the only measurement of this quantity performed to date is that by Senesi {\em et al.} \cite{senesi}, who  studied a much denser system, at higher temperature (2 K).
The value of the kinetic energy per $^3$He atom obtained in our simulation is 23.14(3) K. This result is consistent with the {approximate} ground state estimate of Ref. \onlinecite{whitlock}, namely 23.6 K, obtained for a slightly higher density (0.02509 \AA$^{-3}$) and with a  different pair potential.
%------------
\begin{figure}[!t]
\includegraphics[width=6.0cm]{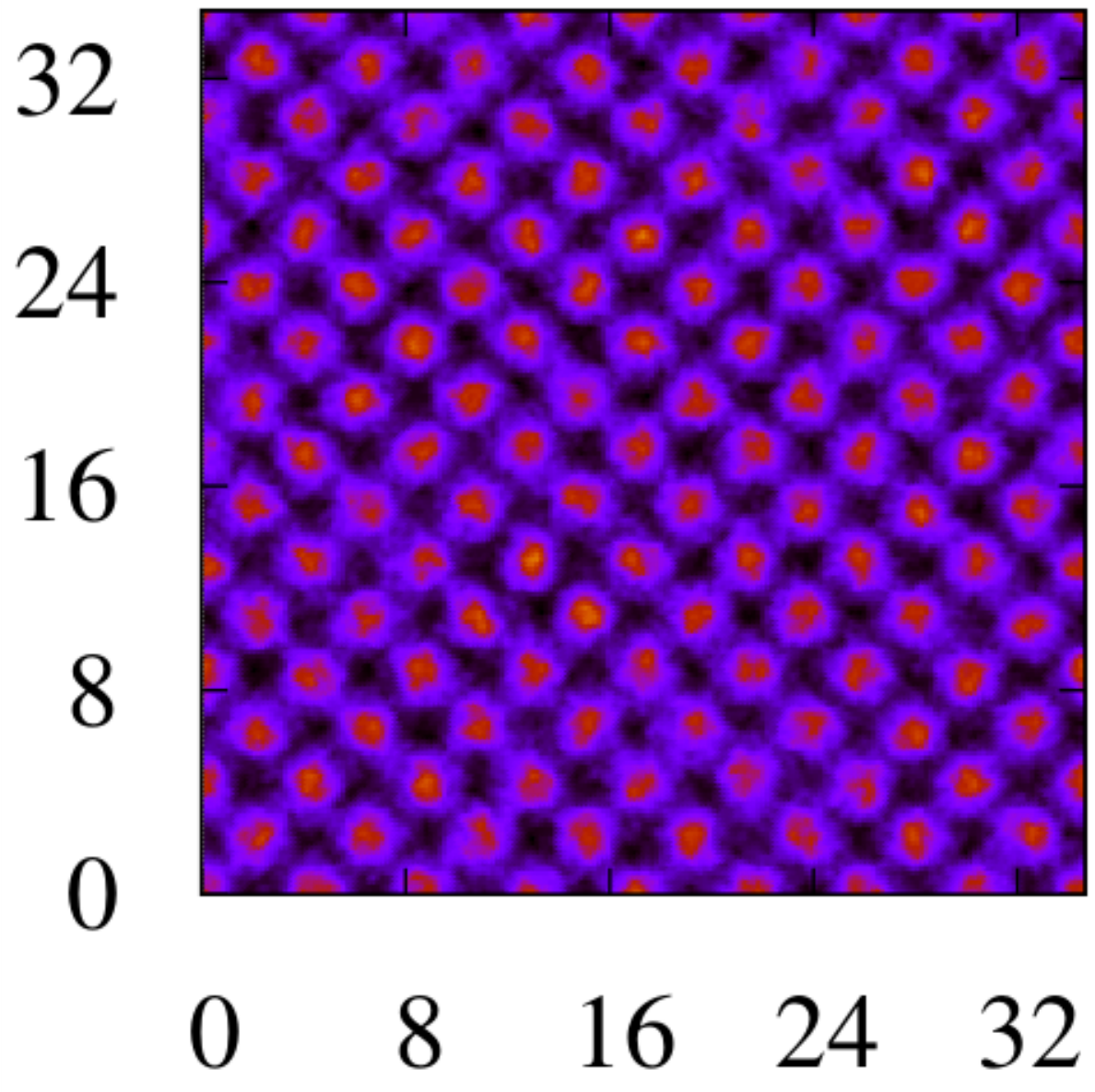} %{figure4}
\caption{{\em Color online.} Snapshot of a many-particle configuration (world lines) for bcc $^3$He, at temperature $T$=0.65 K, density $\rho=0.02458$ \AA$^{-3}$. View is along one of the three main (equivalent) crystallographic directions. The total number of particles is 1024. Only the traces of the 128 in two adjacent planes are visible.
Distances are in \AA.}
\label{f4}
\end{figure}
%---------------
%------------
\begin{figure}[!h]
\includegraphics[width=8.0cm]{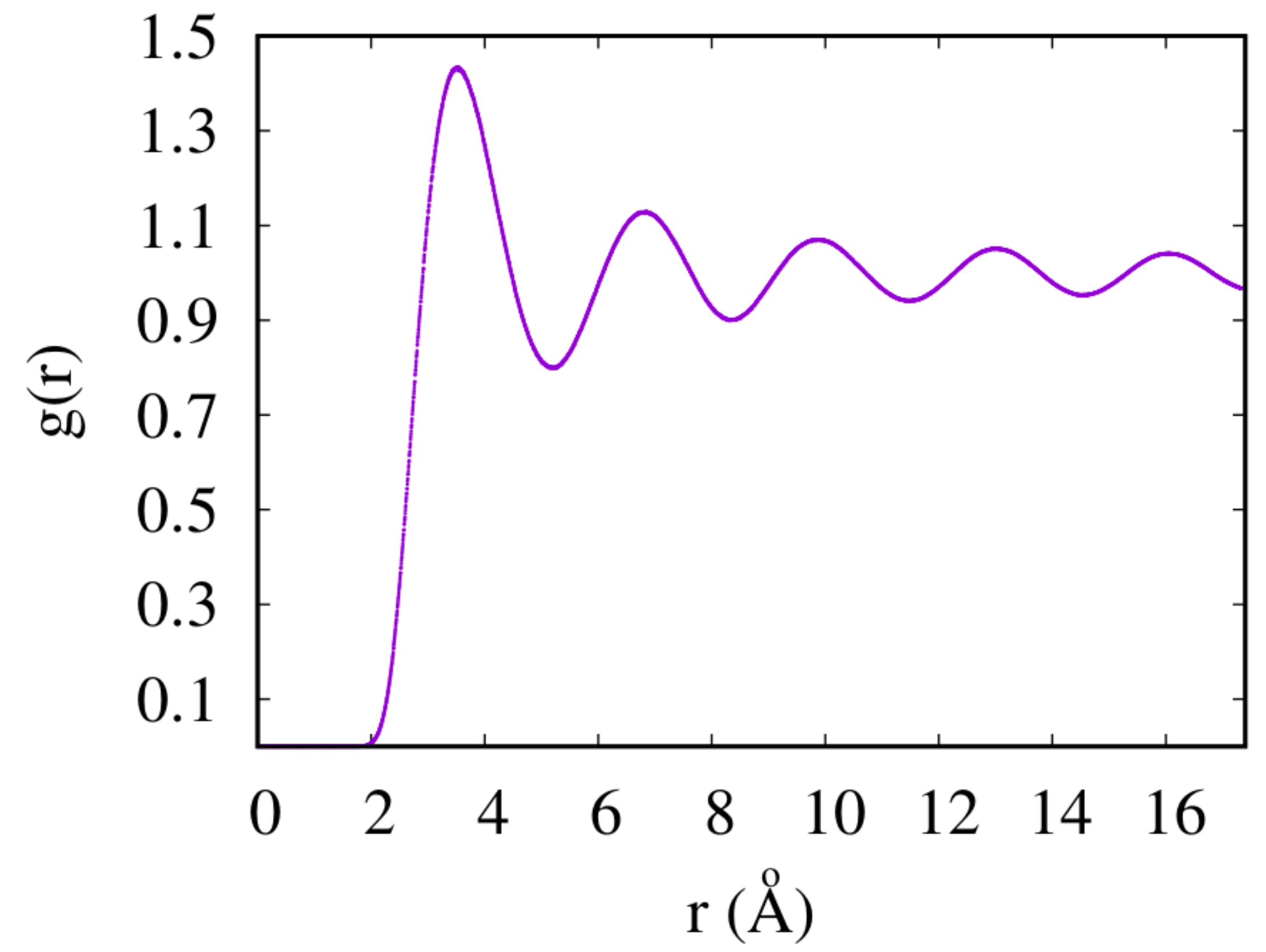} %{figure4}
\caption{{\em Color online.} Pair correlation function $g(r)$ for bcc $^3$He at $T$=0.65 K and density $\rho=0.02458$ \AA$^{-3}$, computed for a system of
$N$=1024 atoms.}
\label{f5}
\end{figure}
%---------------
On comparing the configurational snapshot shown in Fig. \ref{f4} with the corresponding one in Fig. \ref{f1}, one can visually appreciate an important qualitative difference between the $^3$He and $^4$He solids, namely the significantly greater delocalization of the $^3$He atoms. This is quantitatively reflected in the pair correlation function shown in Fig. \ref{f5}. Obvious are the considerably lower height of the first peak ($\sim 1.435$ versus $\sim$ 1.575 for $^4$He), as well as the much weaker secondary oscillations. 
\\ \indent 
All of this points to a significantly more quantal crystal than the $^4$He discussed above, as expected given the lower density and  atomic mass, and as confirmed also by the computed 
value of the parameter $\chi$, which is worth $\sim$ 0.87, i.e., substantially higher than in $^4$He, showing the degree to which $^3$He is the most quantum-mechanical of all naturally occurring crystals. Consequently, one also expects more pronounced atomic displacements away from lattice sites.
%------------
\begin{figure}[!h]
\includegraphics[width=8.0cm]{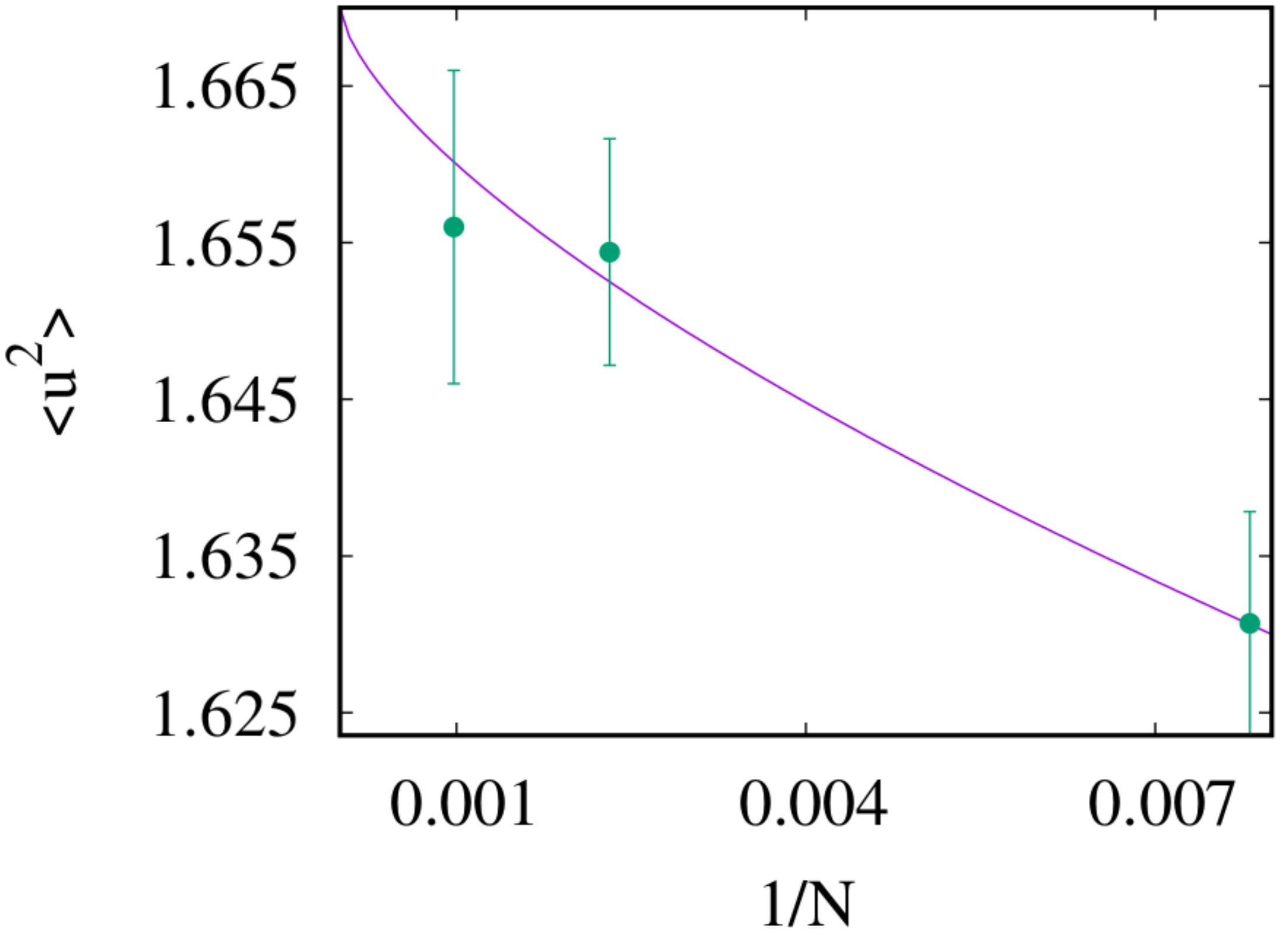} %{figure4}
\caption{{\em Color online.} Mean square atomic displacements (in \AA$^2$) away from lattice sites in bcc $^3$He at the
melting temperature $T$=0.65 K and density $\rho=0.02458$ \AA$^{-3}$. Shown here are the results for the three system sizes simulated (see text). Curve through the points is a fit to the data  based on the expression 
$\alpha+\beta N^{-2/3}$.}
\label{f6}
\end{figure}
%---------------
\\ \indent
Fig. \ref{f6} shows the estimates for $\langle u^2\rangle$ for bcc $^3$He, at temperature $T$=0.65 K and density $\rho=0.02458$ \AA$^{-3}$, for the same three system sizes considered for $^4$He. Within the statistical uncertainty of the calculation, the estimate obtained on a system comprising $N=1024$ atoms is indistinguishable from that for $N=432$. The weak dependence on $N$ reflects the low temperature at which the calculation is carried out. Indeed, we consistently observed that the bulk of the size dependence of $\langle u^2\rangle$ comes from  the (weak) centroid contribution $\langle u^2\rangle_T$, while 
$\langle u^2\rangle_Q$ is essentially independent of $N$.
This is in qualitative agreement with the findings of Ref. \onlinecite{draeger}.
\\ \indent
The same fitting procedure utilized above for $^4$He yields a value of $\langle u^2\rangle$ extrapolated to the thermodynamic limit of $1.67(1)$ \AA$^2$. The extrapolated value is consistent, within statistical errors, with the estimates obtained for $N=432, 1024$ atoms, and indeed it is largely independent on the functional fitting form utilized. The resulting Lindemann ratio is $\delta=0.344(1)$,  somewhat {lower} than the existing theoretical estimate of $\sim$ 0.370 based on time-dependent Hartree \cite{dewette} and self-consistent phonon theory \cite{hernadi}; these calculations are both approximate, and based on different pair potentials than the one utilized here (effects of Fermi statistics of the $^3$He atoms are also not included, as in the calculation carried out here).
\\ \indent
As mentioned above, we are not aware of any measurement of $\langle u^2\rangle$ for bcc $^3$He at the physical conditions considered in this work. On the one hand, the microscopic model utilized here should be reasonably expected to afford the same accuracy for both isotopes, as the pair-wise interaction is very nearly the same,  the magnitude of
non-adiabatic and nuclear spin coupling effects being negligibly small.
On the other hand, quantum exchanges are comparatively more important in solid $^3$He than in $^4$He, owing to the lighter atomic mass and lower density of the former, and definitely increase atomic delocalization.  It seems therefore that a comparison of  the experimentally measured value of $\langle u^2\rangle$ with that predicted in this calculation, in which quantum exchanged are not included, should provide an important clue as to the validity of 
the suggestion of Ref. \onlinecite{pederiva}.

\subsection{Parahydrogen}
We now illustrate our results for \ph2. As we shall see, the main physical observations are rather different from those made for helium; in particular,  while the physical behaviour of bcc He is dominated by quantum mechanics, that of \ph2 reflects an interplay of quantum and thermal effects.
\\ \indent
We obtained estimates for bulk hcp \ph2 at the constant density $\rho=0.0261$ \AA$^{-3}$, 
in the temperature range 1 K $\le T \le $ 13.8 K, i.e., from what is essentially the ground state \cite{omi} all the way to the melting temperature. It is experimentally known that thermal expansion of a \ph2 crystal at low $T$ is remarkably small \cite{alonso}. In this temperature range, the computed kinetic energy per molecule increases monotonically from a $T\to 0$ value of 70.2(1) K to that of 71.9(1) at the melting temperature. These estimates seem altogether consistent with the most recent experimental measurements of $E_k$ \cite{colognesi}; it should be noted how the variation of this quantity in the range of $T$ explored here is of the order of the typical experimental uncertainty.
\\ \indent
%------------
\begin{figure}[!h]
\includegraphics[width=8.6cm]{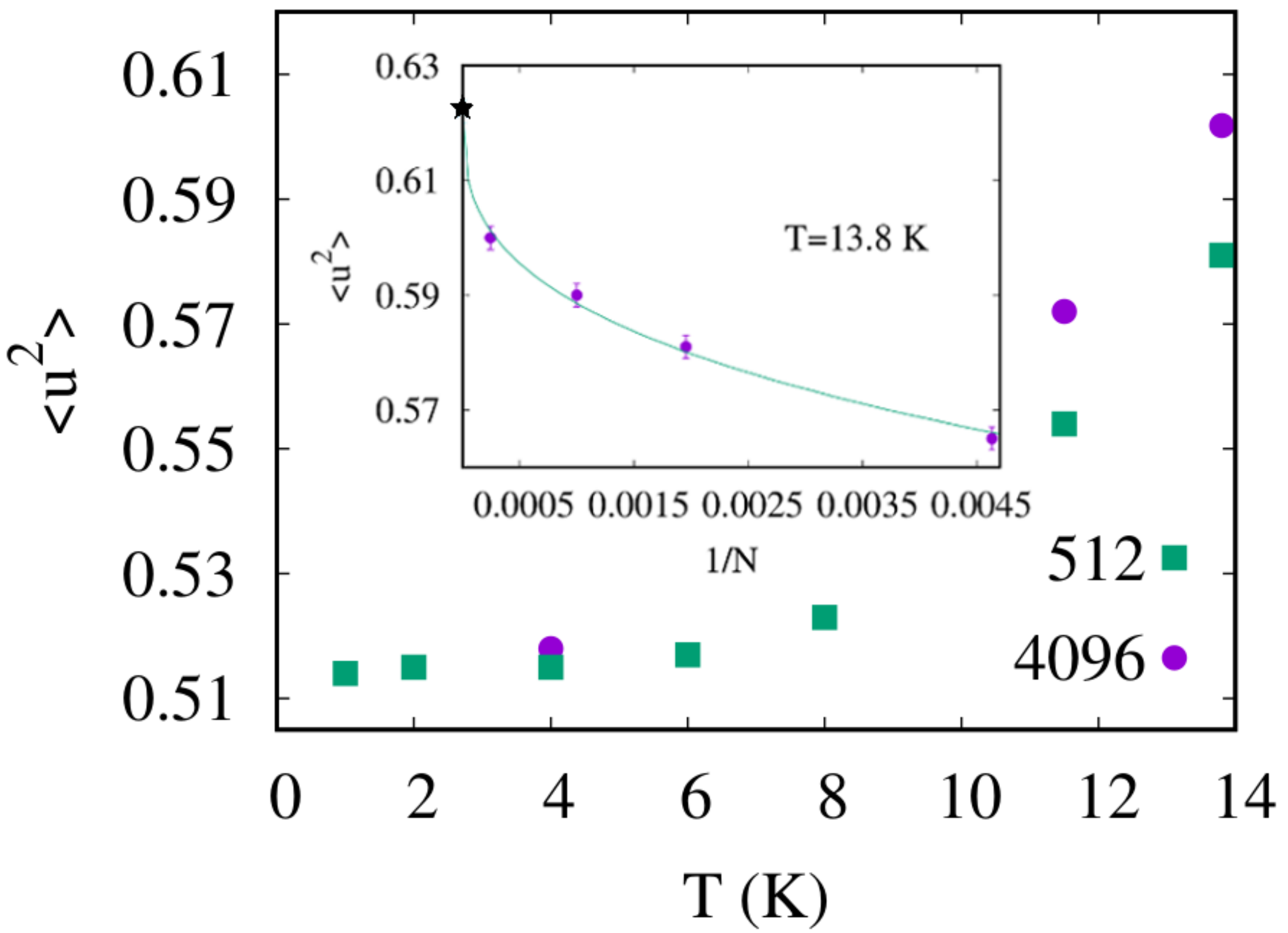} %{figure4}
\caption{{\em Color online.} Mean square atomic displacements (in \AA$^2$) away from lattice sites in hcp \ph2 at different temperatures, at density 
$\rho=0.0261$ \AA$^{-3}$. Shown here are results for two simulated systems, comprising $N$=512 (squares) and $N$=4096 (circles) molecules. When not explicitly shown, statistical errors are smaller of the symbol size. Inset shows the size extrapolation of the results obtained at temperature $T$=13.8 K. Curve through data points is a fit obtained using the expression $\alpha+\beta N^{-1/3}$. Star represents the value extrapolated to the thermodynamic limit. }
\label{f7}
\end{figure}
%---------------
Fig. \ref{f7} shows numerical estimates for $\langle u^2\rangle$ obtained for \ph2 as a function of temperature, for two different system sizes, namely $N$=512 (squares) and $N$=4096 molecules. The first, obvious remark is that the size dependence of the results changes dramatically as the temperature is raised; in particular, at low $T$ (i.e., $T\lesssim 4$ K), where molecular excursions are mostly quantum mechanical, the estimates obtained for the two different system sizes are essentially the same, within  the statistical errors of the calculations. Extrapolation of the results to the thermodynamic limit can be carried out just like for solid He, i.e., the data obtained on different system sizes can be fitted with the functional expression (\ref{fitting}), with the same value of $\gamma$ used for He, namely $\gamma=2/3$.
\\ \indent
On the other hand, on approaching the melting temperature the dependence of the results on system sizes becomes much more significant, and, as observed also in Ref. \onlinecite{draeger}, extrapolation to the thermodynamic limit requires that the parameter $\gamma$ in (\ref{fitting}) be set to 1/3, which is consistent with
excursions away from lattice sites being mainly thermally driven \cite{draeger}. The inset of Fig. \ref{f7} shows the extrapolation to the thermodynamic limit of the estimates for $\langle u^2\rangle$ at the melting temperature $T$=13.8 K. The extrapolated value is 0.622 \AA$^2$, corresponding to a Lindemann ratio equal to
0.208 (nearest neighbor distance is 3.784 \AA), certainly significantly above the value  ($\sim 0.1$) typically associated with thermal melting, confirming that quantum fluctuations play an important, though not dominant role in the behaviour of a \ph2 crystal at melting. Indeed, the value of the parameter $\chi$ introduced above is remarkably close to 0.5 at this temperature, i.e., quantum and thermal contribution to molecular displacements are very nearly equal.
\begin{table}[h]\label{t2}
\centering
  \caption{Calculated mean square displacement of molecules in a \ph2 crystal at various temperatures. The density is $\rho=0.0261$ \AA$^{-3}$. The quoted values of $\langle u^2\rangle$ are extrapolated to the thermodynamic limit.  Also shown is the parameter $\chi\equiv\langle u^2\rangle_Q/\langle u^2\rangle$ for the largest system size considered here ($N$=4096), as well as the Lindemann ratio $\delta$, obtained from $\langle u^2\rangle$ assuming 
a temperature-independent nearest neighbor distance of 3.784 \AA. Last column shows experimental estimates of $\delta$, from Ref. \onlinecite{alonso}.
Statistical uncertainties, in parentheses, are on the last digit.}
  \label{tab:table1}
  \begin{tabularx}{.48\textwidth}{ccccccccccccccccccccc}
\hline\hline\\
  $T$(K)  &&&&& $\langle u^2\rangle$ (\AA$^{2}$) &&&&&$\chi$ &&&&& $\delta$ &&&&&$\delta$ (Ref. \onlinecite{alonso}) \\
    \hline\\
13.8 &&&&& 0.622(2) &&&&& 0.49 &&&&&0.208(1) &&&&& \\
  11.5 &&&&& 0.593(5)&&&&&0.56 &&&&& 0.204(1) &&&&&0.200(2) \\
8.0 &&&&& 0.547(2) &&&&&0.69 &&&&&0.195(1) &&&&& \\
4.0 &&&&& 0.519(2) &&&&&0.84 &&&&&0.190(1) &&&&&0.199(2) \\
\hline
  \end{tabularx}
\end{table}
\\ \indent
We have carried out size extrapolations of $\langle u^2\rangle$ at several other temperatures; we provide some of our estimates in Table II, together with the values of $\chi$ obtained from the largest system simulated here ($N$=4096) \cite{notew}.  Also provided is the Lindemann ratio $\delta$, computed assuming a temperature-independent nearest-neighbor distance. The rightmost column contains experimental estimates of $\delta$ obtained from Ref. \onlinecite {alonso}; specifically, we read off Fig.  2 of Ref. \onlinecite{alonso} the estimates of $u_{rms}$ and divided them by 3.784 \AA.
\\ \indent
Our results, both as illustrated in Fig. \ref{f7} as well as  in Table II, show a clear temperature dependence of the molecular displacements; specifically, the Lindemann ratio increases by $\sim 10$\% from its $T\to 0$ value to that at melting. This is in disagreement with the experimental data reported in Ref. \onlinecite{alonso}, which are quantitatively consistent with our predictions near melting, but feature  no significant temperature dependence, essentially in the entire temperature interval considered here, which extends from essentially ground state all the way to melting. This observation prompted the claim made in Ref. \onlinecite{alonso} that the physics of a \ph2 crystal at saturated vapour pressure is dominated by quantum mechanics.
\\ \indent
The calculation carried out in this work fails to support such a claim; specifically, although there are undoubtedly significant, measurable quantum effects in a crystal of \ph2, even at the melting temperature (just as there are in the liquid phase at freezing \cite{bon09}), as indicated by the relatively high value of the Lindemann ratio, nonetheless thermal effects are significant at high $T$. Indeed, the parameter $\chi$, which is a measure of the relative importance of quantum effects, takes on in \ph2 a value close to 0.5 at melting, as opposed to,  for example, 0.75 of $^4$He (a similar value in \ph2 is observed at $T\sim 7$ K, i.e., one half of the melting temperature).
\\ \indent
The temperature dependence of the mean square molecular displacements observed in our simulations of a \ph2 crystal, is of a magnitude that should lend itself to unambiguous experimental detection. Thus, we have no explanation for the qualitative and quantitative disagreement between our theoretical results and the experimental measurements of Ref. \onlinecite{alonso} at low temperature ($T \lesssim 10$ K). The potential utilized here is the standard intermolecular potential for condensed parahydrogen, and it is difficult to imagine that the discrepancy, which is temperature dependent, could be attributable to it. 
As such a  discrepancy is most significant at low temperature, one might be tempted to ascribe it to the possible effect of quantum exchanges, but these  are  essentially absent in this system, and certainly at the temperatures considered here. 

\section{Conclusions}\label{conclusions}
In this work, we carried out  extensive QMC simulations of the bcc solid phase of the two isotopes of helium, as well as of the hcp phase of \ph2 at low temperature, based on a microscopic model only including pair-wise interactions. In all of the calculations, we regarded particles as {\em distinguishable}, i.e., quantum statistics was ignored. It is worth noting that a recent QMC calculation of the Lindemann ratio for bcc $^4$He including quantum-mechanical exchanges (Ref. \onlinecite{rota}) yielded a result consistent with the one reported here, lending some validation to the hypothesis that exchanges are indeed negligible. The microscopic model affords generally satisfactory to excellent agreement with measured  thermodynamic properties (e.g., the pressure).
\\ \indent
We computed for all the systems considered the mean square displacement of particles away from lattice sites, and obtained estimates in excellent agreement with experiment for bcc $^4$He; the estimate obtained for the
Lindemann ratio in bcc $^3$He   is  about 6.5\% below existing  theoretical predictions. This new result should supersede those predictions, since it is based on a more robust methodology and accurate pair potential. \\ \indent
The microscopic Hamiltonian utiized is the same for both isotopes, and is expected to afford comparable accuracy; it was suggested in Ref. \onlinecite{pederiva}, however, that quantum exchanges significantly enhance atomic mobility in a $^3$He crystal, and thus must be explicitly included in a calculation, if accurate agreement with experiment is sought. 
This would point to a remarkable difference between the roles of Fermi and Bose statistics {\em vis-\`a-vis} crystallization; for, while the latter enhances the stability of the (superfluid) liquid phase \cite{munich12},  the former promotes the formation of a crystal in which particles enjoy much greater mobility than the neglect of statistics would lead one to expect. We did not attempt to perform any simulation explicitly including Fermi statistics in this work, due to the presence of the above-mentioned ``sign" problem. It is not {\em a priori} obvious that such a calculation would be unfeasible, as exchanges  are likely to be still relatively infrequent, and therefore the signal-to-noise ratio may remain high enough to allow one to collect sufficient statistics with reasonable computational effort.  We postpone this project until the future.\\
 \indent
In general, we found that the physics of these two crystal at melting is largely dominated by quantum mechanics (zero-point motion). Indeed, we observed this to remain the case even in the hcp phase of $^4$He at higher pressure. For example, simulation of  hcp $^4$He at a pressure of approximately 40 bars (density $\rho=0.0312$ \AA$^{-3}$), at the melting temperature $T$=2.5 K (using a number of particles $N$=512)  yields a value of the parameter $\chi$ defined above close to 0.7, i.e., most of the contribution to atomic displacements comes from quantum mechanics. On further pressurizing the hcp crystal, the value of $\chi$ at melting reduces progressively, i.e., the behaviour of the system approaches that of a classical solid; however,
effects of quantum mechanics remain significant even at a relatively high pressure. For example, close to 2.4 kbars at the melting temperature $T$=24 K ($\rho=0.05735$ \AA$^{-3}$), $\chi$ is found to be  $\sim$ 0.4.
\\ \indent
Our calculated molecular displacements in solid \ph2 display a clear dependence on temperature, in disagreement with recent experimental measurements. The cause of this discrepancy is unknown at this time. It is found that, although quantum mechanical effect are important, the behaviour of this system near melting 
shows both quantum and classical aspects, both roughly equally important in magnitude.

\section*{Acknowledgments}
This work was supported in part by the Natural Science and Engineering Research Council of Canada. Computing support from Westgrid is gratefully acknowledged.

\noindent
$^\star$ Permanent address: Department of Physics and Physical Oceanography, Memorial
 University of Newfoundland, St. John's,  Newfoundland A1B 3X7, Canada.


\begin{thebibliography}{99}
\bibitem{note0} We adopt here a system of units where the Boltzmann constant $k_B$=1.
\bibitem{noteaa}
 Whenever a new many-particle configuration is sampled,  a rigid shift of the positions of all particles is performed before proceeding with the calculation of  (\ref{tt}),  in order to ensure that  the  average center-of-mass position along the path to be the same for all the many-particle configurations in the statistical sample.
\bibitem{notew}
It should be noted that the value of $\chi$ is very nearly size independent. For example, a simulation at $T$=13.8 K with $N$=216 yields $\chi=0.52$.
\end{thebibliography}
\end{document}